\documentclass[iop]{emulateapj}
\usepackage{natbib}
\usepackage{epstopdf}

\begin{document}

\title{Detection of N$_2$D$^+$ in a protoplanetary disk}

\author{Jane Huang and Karin I. \"Oberg}
\affil{Harvard-Smithsonian Center for Astrophysics, 60 Garden St., Cambridge, MA 02138}

\begin{abstract}
Observations of deuterium fractionation in the solar system, and in interstellar and circumstellar material, are commonly used to constrain the formation environment of volatiles. Toward protoplanetary disks, this approach has been limited by the small number of detected deuterated molecules, i.e. DCO$^+$ and DCN. Based on ALMA Cycle 2 observations toward the disk around the T Tauri star AS 209, we report the first detection of N$_2$D$^+$ (J=3-2) in a protoplanetary disk. These data are used together with  previous Submillimeter Array observations of N$_2$H$^+$ (J=3-2) to estimate a disk-averaged D/H ratio of 0.3--0.5, an order of magnitude higher than disk-averaged ratios previously derived for DCN/HCN and DCO$^+$/HCO$^+$ around other young stars. The high fractionation in N$_2$H$^+$ is consistent with model predictions. The presence of abundant N$_2$D$^+$ toward AS 209 also suggests that N$_2$D$^+$ and the N$_2$D$^+$/N$_2$H$^+$ ratio can be developed into effective probes of deuterium chemistry, kinematics, and ionization processes outside the CO snowline of disks.   
\end{abstract}

\section{Introduction}
The gas and dust disks surrounding young stars are the formation sites of planets. Obtaining a chemical inventory of these disks is key to developing effective molecular probes of disk ionization, temperature, and other characteristics that set planet formation efficiencies. Applying such probes to disk studies can illuminate the processes that led to the present-day composition of solar system bodies and exoplanets. 

The thermal histories of solar system, protoplanetary disk and interstellar material are often probed via measurements of deuterated isotopologues of common molecules (e.g. \citealt{1999ApJ...526..314A, 2011ARA&A..49..471M}). Molecular D/H ratios in cold interstellar and circumstellar environments can be several orders of magnitude greater than the cosmic D/H abundance  of $\sim 10^{-5}$ (e.g., \citealt{2005ApJ...619..379C, 2008ApJ...681.1396Q,2012ApJ...749..162O, 2015A&A...574A.137T}). The increase in the D/H ratio of specific molecules, i.e. the deuterium fractionation, is a consequence of the lower zero-point energies of deuterated molecules compared to their non-deuterated isotopologues. The resulting small energy barrier for a molecule to exchange a deuterium atom for a hydrogen atom favors deuterium fractionation in cold environments. High D/H ratios in the ISM help to identify regions that are currently cold, e.g. pre-stellar cores, and to infer a low-temperature history for material in warmer regions.  In the solar system, measurements of D/H ratios in comets set constraints upon the temperature (and therefore the location) at which they formed in the Solar nebula (e.g. \citealt{2011Natur.478..218H, 2015Sci...347A.387A}), and the high D/H ratio of terrestrial water indicates inheritance from the cold parent molecular cloud \citep{2014Sci...345.1590C}. 

Although more than 30 deuterated molecules have been detected throughout various pre-stellar and protostellar environments (see \citealt{2003SSRv..106...61R}, \citealt{2007prpl.conf...47C} and references therein), only two of them, DCO$^+$ and DCN, had been detected in protoplanetary disks \citep{2003A&A...400L...1V, 2008ApJ...681.1396Q}. A third deuterated molecule that has been predicted to be abundant in disks is N$_2$D$^+$ \citep{2007ApJ...660..441W}.
 N$_2$D$^+$ forms via the reaction \citep{1984ApJ...287L..47D}
\begin{equation}
\textup{H}_2\textup{D}^+ +\textup{ N}_2 \rightarrow \textup{H}_2 + \textup{ N}_2\textup{D}^+.
\end{equation}
The formation of H$_2$D$^+$ compared to H$_3^+$ is enhanced at temperatures below 50 K \citep{1982A&A...111...76H, 2014prpl.conf..859C}. This leads to an increase in the N$_2$D$^+$/N$_2$H$^+$ ratio in cold environments. Because H$_2$D$^+$ is efficiently destroyed by gas-phase CO, abundant N$_2$D$^+$, co-existing with its non-deuterated isotopologue N$_2$H$^+$, is only expected outside the CO snowline, where CO freezes out on grains \citep{2013Sci...341..630Q}. Together with N$_2$H$^+$, N$_2$D$^+$ may be the best and perhaps only tracer of the chemistry and ionization of the midplane in the outer disk region \citep{2014ApJ...794..123C}, the proposed formation site of Kuiper Belt objects, comets, Uranus, and Neptune \citep{2013Sci...341..630Q,2014ApJ...793....9A}. 

As part of an ALMA Cycle 2 disk survey of deuterated molecules, we searched for  N$_2$D$^+$ toward AS 209, a 1.6 Myr old K5 T Tauri star in $\rho$ Ophiuchi located 130$\pm$ 50 pc away \citep{1988cels.book.....H,2007A&A...474..653V}. AS 209 is surrounded by a large and gas-rich disk, extending out to  340 AU in CO emission \citep{2009ApJ...700.1502A, 2011ApJ...734...98O}. Molecules previously detected in the disk include CO, HCO$^+$, DCO$^+$, N$_2$H$^+$, HCN, and CN \citep{2011ApJ...734...98O}. In this letter, we present the first detection of N$_2$D$^+$ in a protoplanetary disk.  We use existing Submillimeter Array observations of N$_2$H$^+$ in the same disk to calculate its deuterium fractionation and discuss the implications for future disk studies. 

\section{Observations}

AS 209 (J2000.0 R.A. $16^{\textup{h}}49^{\textup{m}}15^{\textup{s}}.29$, decl. $-14^{\circ}22'08''.6$) was observed 2014 July 2 during ALMA Cycle 2 with 21 minutes of on-source integration time. Thirty-four 12 m antennae were used for the observations, with a maximum baseline of 650.3 m. Thirteen spectral windows (SPWs) were observed in Band 6, each with 61.04 kHz resolution. The N$_2$D$^+$ J=3-2 line was in SPW 9, centered on 231.32183 GHz, and the $^{13}$CO J=2-1 line (used in this study as a v$_{LSR}$ and spectral profile reference) was in SPW 7, centered  on 220.39868 GHz. The 1.4 mm dust continuum flux was extracted by averaging the channels in five line-free SPWs. Each of these windows was 59 MHz wide. 

ALMA/NAASC staff performed bandpass and phase calibration with the quasar J1733-1304, and the flux was calibrated with Titan. The uncertainty in the absolute flux is $\sim$10$\%$. The calibrated visibilities were deconvolved and CLEANed with the CASA software package (version 4.2.2). The integrated flux density of the 1.4 mm dust continuum is 242$\pm$24 mJy. This was measured by integrating over the area of the disk delimited by a 2$\sigma$ contour, where $\sigma$=0.3 mJy beam$^{-1}$ was the continuum rms measured in a signal-free region of the intensity map. After continuum subtraction in the uv-plane, the data cubes for the SPWs were rebinned to 0.4 km/s. To increase signal to noise, the visibilities for the N$_2$D$^+$ J=3-2 line were tapered to 1$''$.  Because the spatial distribution of the emission changes between different velocity channels, CLEANing for both  N$_2$D$^+$ and  $^{13}$CO was performed using a Keplerian rotation mask based on the emission patterns of the stronger line, $^{13}$CO 2-1. The channel rms for $^{13}$CO 2-1 was 17 mJy beam$^{-1}$, while the rms for N$_2$D$^+$ 3-2 was 8.5 mJy beam$^{-1}$. 

To estimate a D/H ratio for  N$_2$D$^+$/ N$_2$H$^+$, this paper also uses AS 209 data originally published in {\"O}berg et al.'s Submillimeter Array protoplanetary disk survey \citeyearpar{2011ApJ...734...98O}. The 1.4 mm dust continuum flux density from their dataset (also obtained by integrating over the area of the disk encompassed by the  2$\sigma$ contour) is 219 $\pm$ 22 mJy. Thus, within the uncertainties, the flux density from the ALMA observations is consistent with that of the SMA observations.

\section{Results}

\begin{figure}[htp]
\epsscale{1}
\plotone{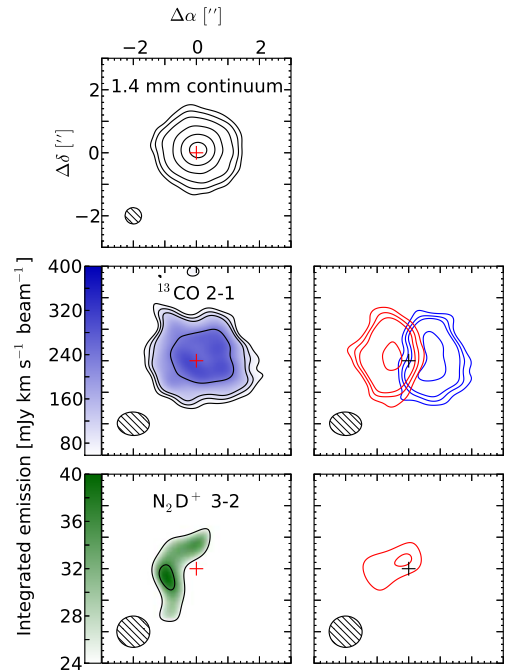}
\caption{Continuum and line emission maps toward AS 209. The topmost panel shows the 1.4 mm dust continuum emission, 
with contours at [5, 10, 20, 40, 80, 160]$\sigma$, where $\sigma$ = 0.3 mJy beam$^{-1}$.  Integrated intensity maps (summed from 2.6 to 7.0 km s$^{-1}$)  with contours at [2, 3, 4, 8]$\sigma$ are shown in the left column for $^{13}$CO 2-1 , with $\sigma$ = 29 mJy km s$^{-1}$  beam$^{-1}$, and N$_2$D$^+$  3-2, with $\sigma$ = 12 mJy km s$^{-1}$  beam$^{-1}$. The corresponding panels in the right column show the integrated intensities for the blueshifted (2.6 to 4.6 km s$^{-1}$) and redshifted (5.0 to 7.0 km s$^{-1}$) emission. The pointing center in each panel is marked by a cross, with position offsets in arcseconds. Synthesized beams are drawn in the lower left of each panel. 
\label{fig1}}
\end{figure}

Figure \ref{fig1} shows the AS 209 1.4 mm dust continuum, integrated intensity maps (spanning channels from 2.6 to 7.0 km s$^{-1}$), and blue and redshifted disk emission maps for the N$_2$D$^+$ 3-2 and $^{13}$CO 2-1 lines. The CO emission is symmetric, whereas most of the emission above the 2$\sigma$ level for N$_2$D$^+$ 3-2 is east of the phase center. The peak intensity of N$_2$D$^+$ is 3.3$\sigma$, where $\sigma$=12 mJy beam$^{-1}$ km s$^{-1}$ is the rms measured in a signal-free portion of the integrated intensity map. The velocity field of $^{13}$CO 2-1 in Fig. \ref{fig1} is well-resolved and consistent with the Keplerian rotation of a disk. The redshifted N$_2$D$^+$ emission is consistent with the redshifted $^{13}$CO emission, but no blueshifted emission is observed for N$_2$D$^+$ above the 2$\sigma$ level. This is spatially consistent with the asymmetry observed in the total integrated intensity map. While the morphologies of the contours differ between the total integrated intensity maps and 
the redshifted emission maps for N$_2$D$^+$ due to the differing noise contributions for the different velocity ranges, the mean intensity measured from the integrated intensity map is within 5$\%$ of the mean intensity measured over the redshifted channels only, confirming that most of the emission is redshifted.

\begin{figure}[htp]
\epsscale{1.0}
\plotone{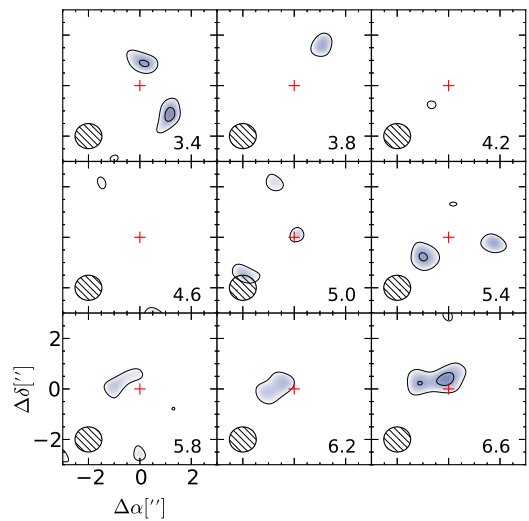}
\caption{Velocity channel maps of N$_2$D$^+$ 3-2 emission toward AS 209, binned by 0.4 km s$^{-1}.$ 2$\sigma$ and 3$\sigma$ contours are drawn, with $\sigma$ = 8.5 mJy beam$^{-1}$. Channel velocities [km s$^{-1}$] are listed in the lower right corner of each panel. The pointing center is denoted by a cross, with offsets from the pointing center indicated in arcseconds on the axes. Synthesized beams are drawn in the lower left of each panel.  \label{fig2}}
\end{figure}

Velocity channel maps for N$_2$D$^+$ 3-2 are shown in Fig. \ref{fig2}. Line flux at $>$3$\sigma$ level is detected near the phase center in three channels, and at the 2$\sigma$ level in another three channels. The progression of emission from west to east of the pointing center is consistent with disk rotation, confirming the disk origin of the N$_2$D$^+$ emission shown in Fig. \ref{fig1}. We note that although blueshifted emission is visible in the velocity channel maps, it is not apparent in the integrated intensity map in Fig. \ref{fig1} above the 2$\sigma$ level because the map is integrated across a velocity range that includes a number of channels without blueshifted emission. 

\begin{deluxetable*}{cccccc}
\tablecolumns{6}
\tablewidth{0pc}
\tablecaption{Summary of Line Observations\label{Table1}}
\tablehead{
\colhead{Transition} & \colhead{Rest Frequency} &\colhead{ E$_u$} &\colhead{ Beam}&\colhead{ Integrated Flux} &\colhead{ Channel rms}\\
\colhead{}& \colhead{(GHz)} & \colhead{(K)} &\colhead{}& \colhead{( mJy km s$^{-1}$) }&\colhead{( mJy beam$^{-1}$})}
\startdata
$^{13}$CO J=2-1\tablenotemark{a} & 220.39868 & 15.9 & 1.$''$02$\times$ 0.$''$74 (87$^{\circ}$.18) &1940[37]&17\\
 N$_2$D$^+$ J=3-2\tablenotemark{a} & 231.32183 & 22.2 &1.$''$05$\times$ 0.$''$99 (87$^{\circ}$.83)&80[18] & 8.5\\
 N$_2$H$^+$ J=3-2\tablenotemark{b} & 279.51173 & 26.8 &3.$''$74$\times$ 2.$''$50 (19$^{\circ}$.26)&597[78]&80  \\

\enddata
\tablenotetext{a}{From ALMA observations.}
\tablenotetext{b} {From data originally published in \citet{2011ApJ...734...98O}. }
\end{deluxetable*}

\begin{figure}[htp]
\epsscale{0.7}
\plotone{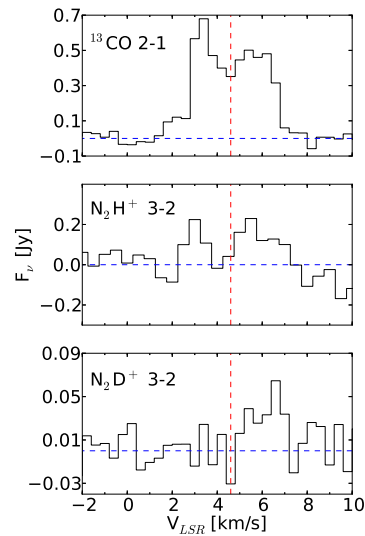}
\caption{Spectra of the $^{13}$CO 2-1 and N$_2$D$^+$  3-2 transitions (extracted with Keplerian masks) toward AS 209, as well as a spectrum of N$_2$H$^+$ 3-2 toward AS 209 from Submillimeter Array data published in \citet{2011ApJ...734...98O}. The v$_{LSR}$ is marked by the vertical dashed lines.
\label{fig3}}
\end{figure}

Figure \ref{fig3} displays spectra for the N$_2$D$^+$ 3-2 and $^{13}$CO 2-1 lines, extracted using the Keplerian CLEAN masks, 
together with the N$_2$H$^+$ 3-2 SMA spectrum. The double-peaked profile of $^{13}$CO 2-1 is typical of a rotating disk. The spectrum of N$_2$H$^+$ 3-2, though with a lower signal-to-noise ratio (SNR), is also consistent with a double-peaked profile. As with the integrated intensity maps, the emission for N$_2$D$^+$ is asymmetric, with only the redshifted peak clearly visible. However, the position of the N$_2$D$^+$ peak is consistent with the positions of the redshifted peaks for N$_2$H$^+$ and $^{13}$CO. The hyperfine structure is not resolved for N$_2$D$^+$. 

The integrated fluxes are given in Table \ref{Table1}. The uncertainty on the integrated flux was calculated using the formula $\sigma = \textup{rms} \times \textup{FWHM}/\sqrt{n_{ch}}$, where the rms is measured from a featureless portion of the spectrum for each transition, the FWHM is measured from the $^{13}$CO 2-1 line, and $n_{ch}$ is the number of channels spanning the FWHM.

Using the  N$_2$H$^+$ line data from \citet{2011ApJ...734...98O} in conjunction with the ALMA observations of  N$_2$D$^+$, we estimate a disk-averaged N$_2$D$^+$/N$_2$H$^+$ ratio toward AS 209.  Because the SNR is low for both molecules, their spatial distributions are not well-constrained by the current set of observations. Hence we assume that the molecules are co-spatial in the disk based on their similar formation and destruction chemistry. In particular, the N$_2$D$^+$ and N$_2$H$^+$ distributions are both expected to be regulated by destructive reactions with CO gas, and therefore the degree of freeze-out of CO and N$_2$. The  N$_2$H$^+$ intensity, averaged over a $5'' \times 5''$ circle centered on the continuum peak, is 320$\pm$40 mJy beam$^{-1}$ km s$^{-1}$, while the N$_2$D$^+$ intensity, averaged over a $4'' \times 4''$ circle centered on the continuum peak, is 9$\pm$2 mJy beam$^{-1}$ km s$^{-1}$. In each case, the surface area over which the intensity was averaged was based on the extent of the molecular emission above the 2$\times$rms level and was largely set by the synthesized beam size. 

Column densities were derived from the integrated intensities using the LTE formula from \citet{2003ApJ...590..314R}, in the case where the emission is optically thin and the continuum brightness temperature is assumed to be much lower than the excitation temperature:  
\begin{equation}
N = \frac{2.04\int\Delta I~dv}{\theta_a \theta_b}\frac{Q_{rot}e^{E_u/T_e}}{\nu^3 \langle S_{ij} \mu^2\rangle} \times 10^{20} \textup{ cm}^{-2}.
\end{equation}
$\int\Delta I~dv$ ( Jy beam$^{-1}$ km s$^{-1}$) is the integrated line intensity, $\theta_a$ and $\theta_b$ (arcsec) are the major and minor axes of the ellipse defined by the FWHM of the Gaussian beam, and $T_e$ (K) is the excitation temperature. The molecular parameters, which are the rotational partition function Q$_{rot}$, transition upper state energy E$_u$ (K), transition rest frequency $\nu$ (GHz), line strength $S$, and electric dipole moment $\mu$ (Debye), were all obtained from the Cologne Database for Molecular Spectroscopy \citep{2001A&A...370L..49M,2005JMoSt.742..215M}.  

The excitation temperature is expected to be between 10 and 25 K in the outer regions of the disk where N$_2$D$^+$ and N$_2$H$^+$ reside (e.g. \citealt{2007ApJ...660..441W,2014prpl.conf..859C}). At 10 K, the disk-averaged column density of N$_2$D$^+$  is $2.1\pm 0.5\times 10^{11}$ cm$^{-2}$ and that of  N$_2$H$^+$  is  $6.3\pm 0.8\times 10^{11}$ cm$^{-2}$. At 25 K, the disk-averaged column density of N$_2$D$^+$  is $1.4\pm 0.3 \times 10^{11}$ cm$^{-2}$ and that of  N$_2$H$^+$  is  $3.1\pm  0.4 \times 10^{11}$ cm$^{-2}$. At a given temperature, the dominant source of uncertainty in the LTE approximation for the column density is due to the uncertainty on the intensity measurement. The resulting D/H ratio is estimated to be between 0.3 and 0.5. As long as the N$_2$D$^+$ and  N$_2$H$^+$ are co-spatial, the ratios of their disk-averaged column densities can be estimated with greater confidence than their absolute column densities, which are more sensitive to the assumed surface area of the emission.  

To check the validity of the optically thin LTE approximation for estimating the column densities of N$_2$H$^+$ and N$_2$D$^+$  toward AS 209, column densities for N$_2$H$^+$ were also calculated with RADEX, a 1D non-LTE radiative transfer code \citep{2007A&A...468..627V}. To cover the range of typical disk conditions, we ran a grid of models with kinetic temperatures between 10 and 25 K and an H$_2$ number density (representing the dominant collision partner) of $10^{6}$--$10^{10}$ cm$^{-3}$. The resulting column densities ranged from 1.7 to 7.8$\times 10^{11}$ cm$^{-2}$, similar to the column density range estimated through LTE calculations. 

\section{Discussion}

\subsection{N$_2$H$^+$ deuterium fractionation}

The estimated N$_2$D$^+$/N$_2$H$^+$ disk-averaged abundance ratio of 0.3-0.5 is consistent with the disk models of \citet{2007ApJ...660..441W}, in which the D/H ratio was predicted to range between 0.35 and 7.0 in the outer disk. The predicted range reflects the dependence of deuterium chemistry on desorption from dust grains due to cosmic rays, thermal processes, and UV radiation, which varied between models. \citet{2007ApJ...660..441W} also predicted that the D/H ratio for N$_2$D$^+$ in the outer disk would be one or two orders of magnitude greater than the corresponding ratios for DCN and  DCO$^+$. This is consistent with our finding that the disk-averaged D/H ratio for N$_2$H$^+$  toward AS 209 is at least an order of magnitude higher than the disk-averaged ratios derived for HCN and HCO$^{+}$ toward other disks. A disk-averaged DCN/HCN abundance ratio of 0.017 was derived for TW Hya \citep{2012ApJ...749..162O}. Disk-averaged DCO$^+$/HCO$^+$ abundance ratios have been estimated to be 4$\times 10^{-3}$ toward DM Tau, 0.02 toward HD 163296, and 0.035 toward TW Hya \citep{2003A&A...400L...1V, 2006A&A...448L...5G, 2013A&A...557A.132M}. However, analyses of spatially resolved data for these three disks have found local DCO$^+$/HCO$^+$ ratios rising as high as 0.1-0.3, with radial variations in the abundances due to radial changes in the temperature \citep{2008ApJ...681.1396Q, 2013A&A...557A.132M, 2015A&A...574A.137T}. The local enhancement levels revealed by the spatially resolved data indicates the value of obtaining spatially resolved data for other deuterated lines, such as N$_2$D$^+$. 

The high D/H ratio for N$_2$D$^+$/N$_2$H$^+$ toward AS 209 and the lower DCO$^+$/HCO$^{^+}$ ratios found in other protoplanetary disks suggest that the deuterium chemistry in the outer disk resembles that of dense cores and protostellar envelopes.  A survey of low-mass starless cores found N$_2$D$^+$/N$_2$H$^+$ ratios ranging from \textless .02 to .44 \citep{2005ApJ...619..379C}, and a survey of Class 0 protostars found N$_2$D$^+$/N$_2$H$^+$ ratios ranging from \textless .034 to .27 \citep{2009A&A...493...89E}. For the Class 0 protostars in which both N$_2$D$^+$ and DCO$^+$ have been measured, the N$_2$D$^+$/N$_2$H$^+$ ratio often exceeds the DCO$^+$/ HCO$^+$ ratio by an order of magnitude \citep{2004A&A...416..603J, 2009A&A...493...89E}. Similarly, toward the dense core L1544, Caselli et al. \citeyearpar{2002ApJ...565..344C} found that N$_2$D$^+$/N$_2$H$^+$  was $\sim$0.2, versus a ratio of 0.04 for DCO$^+$/ HCO$^+$. 

Given that models posit similar formation pathways for N$_2$D$^+$ and DCO$^+$, it may be surprising that such different D/H ratios are predicted and observed for the two molecules. This difference arises because gas-phase CO, a parent molecule of DCO$^+$, destroys H$_2$D$^+$, one of the main parent molecules of N$_2$H$^+$ and N$_2$D$^+$ \citep{2013ApJ...765...34Q}. Due to this destruction pathway, N$_2$D$^+$ is only abundant at low temperatures, where deuterium fractionation is especially efficient and gas-phase CO is depleted due to freezeout onto grains. 

\subsection{N$_2$D$^+$ emission asymmetry}

The N$_2$D$^+$ emission toward AS 209 appears to be asymmetric. Higher SNR observations are required to confirm the reality of this asymmetry, but it is worth considering possible causes for this tentative azimuthal N$_2$D$^+$ variation. Previously observed azimuthal variations in disk molecular emission have been ascribed to foreground absorption, projection effects due to disk inclination, and dust traps cooling nearby gas (e.g. \citealt{2011ApJ...734...98O, 2013ApJ...774...16R, 2014ApJ...792L..25V}). 
Foreground absorption is unlikely to cause asymmetric emission for N$_2$D$^+$ due to the molecule's low abundance and high critical density.  For molecular emission originating close to the cold midplane, as is expected for N$_2$D$^+$, projection effects should also play a minimal role \citep{2013ApJ...774...16R}. Asymmetry in the dust structure is an unlikely cause, given the axisymmetric appearance of the 1.4 mm continuum emission shown in Fig. \ref{fig1}. However, because the N$_2$D$^+$ distribution is expected to be especially sensitive to ionization and temperature, possible causes for asymmetry include X-ray flares or azimuthal density gradients in the outer disk (e.g.  \citealt{2013A&A...550L...8B, 2015ApJ...799..204C}). Temperature asymmetries could be probed with D/H ratios derived from high SNR spatially-resolved observations of N$_2$D$^{+}$ and  N$_2$H$^{+}$, while asymmetry in the midplane ionization structure could be evaluated by comparing N$_2$D$^+$ to neutral tracers such as H$_2$CO, another molecule expected to form past the CO snowline \citep{2013ApJ...765...34Q}. 

\subsection{Developing N$_2$D$^+$ as a disk probe}

The similarity between the deuterium fractionation in disks and interstellar cloud cores suggest that analogous to its use as a tracer of CO-depleted cold gas in cores \citep{2002ApJ...565..331C, 2009A&A...493...89E}, N$_2$D$^+$ with N$_2$H$^+$ could be developed into a unique tracer of the cold, CO-depleted outer disk midplane. In addition, measurements of N$_2$D$^+$ and N$_2$H$^+$, together with DCO$^+$ and HCO$^+$, have been used to estimate the ionization fraction of cloud cores \citep{1998ApJ...499...234C, 2002ApJ...565..344C}. This approach could be adapted to evaluate the ionization fraction of disks.   

Spatially resolved observations through ALMA would allow the D/H ratio as a function of radius to be derived, thereby providing a stronger test for chemical theories regarding how N$_2$D$^+$ is formed and destroyed throughout disks. A sensitive survey of  N$_2$D$^+$ and  N$_2$H$^+$ in disks would also be useful to determine the relationship between deuterium fractionation and other properties of the disk, such as mass, age, and ionization processes.

As molecules with abundances anti-correlated with the abundances of CO isotopologues and related species (e.g. HCO$^+$, DCO$^+$), N$_2$D$^+$ and N$_2$H$^+$ complement the current set of common disk tracers. While previously detected disk molecules have been abundant in the inner disk and in the warm molecular layer high above the midplane, N$_2$D$^+$ and N$_2$H$^+$ are suitable for tracing kinematics, ionization, and temperature structure in the outer disk midplane. Although H$_2$D$^+$ is also expected to be abundant in the outer disk midplane, it has not yet been detected in disks and is expected to require very long integration \citep{2014ApJ...794..123C}. Therefore, further observations and modeling of N$_2$D$^+$ and N$_2$H$^+$ are key to developing a fuller picture of how disk characteristics vary spatially. Constraining such properties can clarify how volatiles have been formed, transported, or destroyed in disks, giving rise to the planets we observe now.

\acknowledgments
This paper makes use of the following ALMA data: ADS/JAO.ALMA\#2013.1.00226. ALMA is a partnership of ESO (representing its member states), NSF (USA) and NINS (Japan), together with NRC (Canada) and NSC and ASIAA (Taiwan), in cooperation with the Republic of Chile. The Joint ALMA Observatory is operated by ESO, AUI/NRAO and NAOJ. The National Radio Astronomy Observatory is a facility of the National Science Foundation operated under cooperative agreement by Associated Universities, Inc. We thank Adam Leroy and the NAASC for assistance with calibration and imaging, Ryan Loomis and Viviana Guzman for helpful discussions on the data, and the referee for useful comments on the paper. K. I. O. also acknowledges funding from the Simons Collaboration on the Origins of Life (SCOL), the Alfred P. Sloan Foundation, and the David and Lucile Packard Foundation.


\begin{thebibliography}{}
\expandafter\ifx\csname natexlab\endcsname\relax\def\natexlab#1{#1}\fi

\bibitem[{{Aikawa} \& {Herbst}(1999)}]{1999ApJ...526..314A}
{Aikawa}, Y., \& {Herbst}, E. 1999, \apj, 526, 314

\bibitem[{{Ali-Dib} {et~al.}(2014){Ali-Dib}, {Mousis}, {Petit}, \&
  {Lunine}}]{2014ApJ...793....9A}
{Ali-Dib}, M., {Mousis}, O., {Petit}, J.-M., \& {Lunine}, J.~I. 2014, \apj,
  793, 9

\bibitem[{{Altwegg} {et~al.}(2015){Altwegg}, {Balsiger}, {Bar-Nun},
  {Berthelier}, {Bieler}, {Bochsler}, {Briois}, {Calmonte}, {Combi}, {De
  Keyser}, {Eberhardt}, {Fiethe}, {Fuselier}, {Gasc}, {Gombosi}, {Hansen},
  {H{\"a}ssig}, {J{\"a}ckel}, {Kopp}, {Korth}, {LeRoy}, {Mall}, {Marty},
  {Mousis}, {Neefs}, {Owen}, {R{\`e}me}, {Rubin}, {S{\'e}mon}, {Tzou}, {Waite},
  \& {Wurz}}]{2015Sci...347A.387A}
{Altwegg}, K., {Balsiger}, H., {Bar-Nun}, A., {et~al.} 2015, Science, 347, A387

\bibitem[{{Andrews} {et~al.}(2009){Andrews}, {Wilner}, {Hughes}, {Qi}, \&
  {Dullemond}}]{2009ApJ...700.1502A}
{Andrews}, S.~M., {Wilner}, D.~J., {Hughes}, A.~M., {Qi}, C., \& {Dullemond},
  C.~P. 2009, \apj, 700, 1502

\bibitem[{{Birnstiel} {et~al.}(2013){Birnstiel}, {Dullemond}, \& {Pinilla}}]{2013A&A...550L...8B}
{Birnstiel}, T., {Dullemond}, C.~P.,\& {Pinilla}, P. 2013, \aap, 550, L8

\bibitem[{{Caselli} {et~al.}(1998){Casellii}, {Walmsley}, {Terzieva} \& {Herbst}}]{1998ApJ...499...234C}
{Caselli}, P., {Walmsley}, C.~M., {Terzieva}, R., {et~al.} 1998, \apj, 499, 234


\bibitem[{{Caselli} {et~al.}(2002{\natexlab{a}}){Caselli}, {Walmsley}, {Zucconi}, {Tafalla},
  {Dore}, \& {Myers}}]{2002ApJ...565..331C}
{Caselli}, P., {Walmsley}, C.~M., {Zucconi}, A., {et~al.} 2002{\natexlab{a}}, \apj, 565, 331

\bibitem[{{Caselli} {et~al.}(2002{\natexlab{b}}){Caselli}, {Walmsley}, {Zucconi}, {Tafalla},
  {Dore}, \& {Myers}}]{2002ApJ...565..344C}
{Caselli}, P., {Walmsley}, C.~M., {Zucconi}, A., {et~al.} 2002{\natexlab{b}}, \apj, 565, 344

\bibitem[{{Ceccarelli} {et~al.}(2014){Ceccarelli}, {Caselli},
  {Bockel{\'e}e-Morvan}, {Mousis}, {Pizzarello}, {Robert}, \&
  {Semenov}}]{2014prpl.conf..859C}
{Ceccarelli}, C., {Caselli}, P., {Bockel{\'e}e-Morvan}, D., {et~al.} 2014,
  Protostars and Planets VI, 859

\bibitem[{{Ceccarelli} {et~al.}(2007){Ceccarelli}, {Caselli}, {Herbst},
  {Tielens}, \& {Caux}}]{2007prpl.conf...47C}
{Ceccarelli}, C., {Caselli}, P., {Herbst}, E., {Tielens}, A.~G.~G.~M., \&
  {Caux}, E. 2007, Protostars and Planets V, 47

\bibitem[{{Cleeves} {et~al.}(2014{\natexlab{a}}){Cleeves}, {Bergin}, \&
  {Adams}}]{2014ApJ...794..123C}
{Cleeves}, L.~I., {Bergin}, E.~A., \& {Adams}, F.~C. 2014{\natexlab{a}}, \apj,
  794, 123

\bibitem[{{Cleeves} {et~al.}(2014{\natexlab{b}}){Cleeves}, {Bergin},
  {Alexander}, {Du}, {Graninger}, {{\"O}berg}, \&
  {Harries}}]{2014Sci...345.1590C}
{Cleeves}, L.~I., {Bergin}, E.~A., {Alexander}, C.~M.~O.~., {et~al.}
  2014{\natexlab{b}}, Science, 345, 1590

\bibitem[{{Cleeves} {et~al.}(2015){Cleeves}, {Bergin}, {Qi}, {Adams}, \& 
{{\"O}berg}}]{2015ApJ...799..204C}
{Cleeves}, L.~I., {Bergin}, E.~A.,  {Qi}, C., {Adams}, F.~C. \& {{\"O}berg}, K.~I. 2015, \apj,
  799, 204

\bibitem[{{Crapsi} {et~al.}(2005){Crapsi}, {Caselli}, {Walmsley}, {Myers},
  {Tafalla}, {Lee}, \& {Bourke}}]{2005ApJ...619..379C}
{Crapsi}, A., {Caselli}, P., {Walmsley}, C.~M., {et~al.} 2005, \apj, 619, 379

\bibitem[{{Dalgarno} \& {Lepp}(1984)}]{1984ApJ...287L..47D}
{Dalgarno}, A., \& {Lepp}, S. 1984, \apjl, 287, L47

\bibitem[{{Emprechtinger} {et~al.}(2009){Emprechtinger}, {Caselli}, {Volgenau},
  {Stutzki}, \& {Wiedner}}]{2009A&A...493...89E}
{Emprechtinger}, M., {Caselli}, P., {Volgenau}, N.~H., {Stutzki}, J., \&
  {Wiedner}, M.~C. 2009, \aap, 493, 89

\bibitem[{{Guilloteau} {et~al.}(2006){Guilloteau}, {Pi{\'e}tu}, {Dutrey}, \&
  {Gu{\'e}lin}}]{2006A&A...448L...5G}
{Guilloteau}, S., {Pi{\'e}tu}, V., {Dutrey}, A., \& {Gu{\'e}lin}, M. 2006,
  \aap, 448, L5

\bibitem[{{Hartogh} {et~al.}(2011){Hartogh}, {Lis}, {Bockel{\'e}e-Morvan}, {de
  Val-Borro}, {Biver}, {K{\"u}ppers}, {Emprechtinger}, {Bergin}, {Crovisier},
  {Rengel}, {Moreno}, {Szutowicz}, \& {Blake}}]{2011Natur.478..218H}
{Hartogh}, P., {Lis}, D.~C., {Bockel{\'e}e-Morvan}, D., {et~al.} 2011, \nat,
  478, 218

\bibitem[{{Herbig} \& {Bell}(1988)}]{1988cels.book.....H}
{Herbig}, G.~H., \& {Bell}, K.~R. 1988, {Third Catalog of Emission-Line Stars
  of the Orion Population : 3 : 1988}

\bibitem[{{Herbst}(1982)}]{1982A&A...111...76H}
{Herbst}, E. 1982, \aap, 111, 76

\bibitem[{{J{\o}rgensen} {et~al.}(2004){J{\o}rgensen}, {Sch{\"o}ier}, \& {van
  Dishoeck}}]{2004A&A...416..603J}
{J{\o}rgensen}, J.~K., {Sch{\"o}ier}, F.~L., \& {van Dishoeck}, E.~F. 2004,
  \aap, 416, 603

\bibitem[{{M{\"u}ller} {et~al.}(2005){M{\"u}ller}, {Schl{\"o}der}, {Stutzki},
  \& {Winnewisser}}]{2005JMoSt.742..215M}
{M{\"u}ller}, H.~S.~P., {Schl{\"o}der}, F., {Stutzki}, J., \& {Winnewisser}, G.
  2005, Journal of Molecular Structure, 742, 215

\bibitem[{{M{\"u}ller} {et~al.}(2001){M{\"u}ller}, {Thorwirth}, {Roth}, \&
  {Winnewisser}}]{2001A&A...370L..49M}
{M{\"u}ller}, H.~S.~P., {Thorwirth}, S., {Roth}, D.~A., \& {Winnewisser}, G.
  2001, \aap, 370, L49

\bibitem[{{Mathews} {et~al.}(2013){Mathews}, {Klaasen}, {Juh{\'a}sz}, {Harsono},
  {Chapillon}, {van Dishoeck}, {Espada}, {de Gregorio-Monsalvoi}, {Hales}, {Hogerheijde},
{Mottram}, {Rawlings}, {Takahashi} \&  {Testi}}]{2013A&A...557A.132M}
{Mathews}, G.~S., {Klaassen}, P.~D., {Juh{\'a}sz}, A., {et~al.} 2013, \aap, 557, 132

\bibitem[{{Mumma} \& {Charnley}(2011)}]{2011ARA&A..49..471M}
{Mumma}, M.~J., \& {Charnley}, S.~B. 2011, \araa, 49, 471

\bibitem[{{{\"O}berg} {et~al.}(2011){{\"O}berg}, {Qi}, {Fogel}, {Bergin},
  {Andrews}, {Espaillat}, {Wilner}, {Pascucci}, \&
  {Kastner}}]{2011ApJ...734...98O}
{{\"O}berg}, K.~I., {Qi}, C., {Fogel}, J.~K.~J., {et~al.} 2011, \apj, 734, 98

\bibitem[{{{\"O}berg} {et~al.}(2012){{\"O}berg}, {Qi}, {Wilner}, \&
  {Hogerheijde}}]{2012ApJ...749..162O}
{{\"O}berg}, K.~I., {Qi}, C., {Wilner}, D.~J., \& {Hogerheijde}, M.~R. 2012,
  \apj, 749, 162

\bibitem[{{Qi} {et~al.}(2013{\natexlab{a}}){Qi}, {{\"O}berg}, \&
  {Wilner}}]{2013ApJ...765...34Q}
{Qi}, C., {{\"O}berg}, K.~I., \& {Wilner}, D.~J. 2013{\natexlab{a}}, \apj, 765,
  34
\bibitem[{{Qi} {et~al.}(2013{\natexlab{b}}){Qi}, {{\"O}berg}, {Wilner},
  {D'Alessio}, {Bergin}, {Andrews}, {Blake}, {Hogerheijde}, \& {van
  Dishoeck}}]{2013Sci...341..630Q}
{Qi}, C., {{\"O}berg}, K.~I., {Wilner}, D.~J., {et~al.} 2013{\natexlab{b}},
  Science, 341, 630


\bibitem[{{Qi} {et~al.}(2008){Qi}, {Wilner}, {Aikawa}, {Blake}, \&
  {Hogerheijde}}]{2008ApJ...681.1396Q}
{Qi}, C., {Wilner}, D.~J., {Aikawa}, Y., {Blake}, G.~A., \& {Hogerheijde},
  M.~R. 2008, \apj, 681, 1396


\bibitem[{{Remijan} {et~al.}(2003){Remijan}, {Snyder}, {Friedel}, {Liu}, \&
  {Shah}}]{2003ApJ...590..314R}
{Remijan}, A., {Snyder}, L.~E., {Friedel}, D.~N., {Liu}, S.-Y., \& {Shah},
  R.~Y. 2003, \apj, 590, 314

\bibitem[{{Rosenfeld} {et~al.}(2013){Rosenfeld}, {Andrews}, {Hughes}, {Wilner},
  \& {Qi}}]{2013ApJ...774...16R}
{Rosenfeld}, K.~A., {Andrews}, S.~M., {Hughes}, A.~M., {Wilner}, D.~J., \&
  {Qi}, C. 2013, \apj, 774, 16

\bibitem[{{Roueff} \& {Gerin}(2003)}]{2003SSRv..106...61R}
{Roueff}, E., \& {Gerin}, M. 2003, \ssr, 106, 61

\bibitem[{{Teague} {et~al.}(2015){Teague}, {Semenov}, {Guilloteau}, {Henning},
  {Dutrey}, {Wakelam}, {Chapillon}, \& {Pietu}}]{2015A&A...574A.137T}
{Teague}, R., {Semenov}, D., {Guilloteau}, S., {et~al.} 2015, \aap, 574, A137

\bibitem[{{van der Plas} {et~al.}(2014){van der Plas}, {Casassus},
  {M{\'e}nard}, {Perez}, {Thi}, {Pinte}, \&
  {Christiaens}}]{2014ApJ...792L..25V}
{van der Plas}, G., {Casassus}, S., {M{\'e}nard}, F., {et~al.} 2014, \apjl,
  792, L25

\bibitem[{{van der Tak} {et~al.}(2007){van der Tak}, {Black}, {Sch{\"o}ier},
  {Jansen}, \& {van Dishoeck}}]{2007A&A...468..627V}
{van der Tak}, F.~F.~S., {Black}, J.~H., {Sch{\"o}ier}, F.~L., {Jansen}, D.~J.,
  \& {van Dishoeck}, E.~F. 2007, \aap, 468, 627

\bibitem[{{van Dishoeck} {et~al.}(2003){van Dishoeck}, {Thi}, \& {van
  Zadelhoff}}]{2003A&A...400L...1V}
{van Dishoeck}, E.~F., {Thi}, W.-F., \& {van Zadelhoff}, G.-J. 2003, \aap, 400,
  L1

\bibitem[{{van Leeuwen}(2007)}]{2007A&A...474..653V}
{van Leeuwen}, F. 2007, \aap, 474, 653

\bibitem[{{Willacy}(2007)}]{2007ApJ...660..441W}
{Willacy}, K. 2007, \apj, 660, 441

\end{thebibliography}
\end{document}